**Characterization of protein complexes using chemical cross-linking coupled electrospray mass spectrometry**

Timothy D. Cummins* and Gopal P. Sapkota*

Medical Research Council Protein Phosphorylation and Ubiquitylation Unit, University of Dundee, Dow St., Dundee DD1 5EH, UK

*Address correspondence to t.cummins@dundee.ac.uk and g.sapkota@dundee.ac.uk

## Abstract

Identification and characterization of large protein complexes is a mainstay of biochemical toolboxes. Utilization of cross-linking chemicals can facilitate the capture and identification of transient or weak interactions of a transient nature (*1, 2*). Here we describe a detailed methodology for cell culture based proteomic approach. We describe the generation of cells stably expressing green fluorescent protein (GFP)-tagged proteins under the tetracycline-inducible promoter and subsequent proteomic analysis of GFP-interacting proteins. We include a list of proteins that were identified as interactors of GFP.

## 1. Introduction

Ectopic introduction of an inducible GFP-tagged protein in cells is a useful technique for characterization of protein-protein interactions. Introduction of a pcDNA-5/FRT/TO$^{TM}$ expression vector (Invitrogen) GFP-tagged gene with a pOG44 Flip-recombinase expression plasmid into Flp-IN$^{TM}$ T-REx$^{TM}$ 293 cells (Invitrogen) will integrate the tagged gene into a single pre-designated FRT site. Following integration of the GFP-tagged gene, cells are treated with hygromycin to select for cells that have successfully integrated the pCDNA 5/FRT/TO vector containing the gene of interest.

Cells are then maintained in growth medium supplemented with hygromycin. The stable integration of GFP-tagged genes in T-Rex$^{TM}$ 293 cells allows for real-time detection of protein expression and troubleshooting for appropriate induction time and dose of doxycycline or tetracycline. Once stable cells are established, a time-course of doxycyline or tetracycline can be performed to determine appropriate duration to achieve desired levels of protein expression (Figure 1).

Purification of GFP or GFP-tagged proteins in complex is performed in concert with the addition of dithiobis succinimidyl propionate (DSP) during lysis. DSP has dual NHS esters (see DSP structure in Figure 2), thus targeting primary amines in protein complexes and forming a conjugated network of proteins (*3, 4*). Lysis buffer is gently supplemented with DSP prior to lysis in a dropwise fashion to minimize precipitation, appropriate amount of lysis buffer is then added to cells and allows DSP to interact with primary amines in proteins and complexes of the proteins. Figure 3 illustrates a graphical representation of the pull-down workflow and protein identification by mass spectrometry.

## 2. Materials

1. TREx 293 cell lines are available from Invitrogen (R780-07). Maintain cells in DMEM supplemented with 10% fetal bovine serum (FBS, Hyclone), 2 mM L-glutamine (Gibco), 100 U/ml pencillin, 200 μg/ml streptomycin, 15 μg/ml blasticidin and 100 μg/ml zeocin.
2. Recombinase Flp-IN pOG44 (V6005-20) and pCDNA5 FRT/TO (V6520-20) can be purchased from Invitrogen.
3. Clone individual genes of interest into pCDNA5 FRT/TO using convenient

multiple cloning sites.

4. Lysis buffer is prepared as 40 mM HEPES pH 7.4, 120 mM NaCl, 1 mM EDTA, 10 mM sodium pyrophosphate, 50 mM NaF, 1.5 mM sodium orthovanadate, 1% Triton, 1 tablet complete protease inhibitor cocktail per 25 ml (*5*).
5. Prepare dithiobis succinimidyl propionate (DSP, Pierce) 200 mg/ml in dimethyl sulfoxide (prepare fresh). DSP forms an NHS ester with primary amines and is cleaved upon reduction with dithiothreitol (DTT). This property allows for purification of conjugated DSP-protein complexes and release upon addition of the reducing agent in the final steps just prior to performing SDS-PAGE.
6. Equilibrate GFP-trap beads (ChromoTek) in lysis buffer (above). Finally, prepare 50% slurry.
7. 1.5 mm NuPAGE Bis-Tris 10% precast gels (Invitrogen).
8. MOPS SDS-PAGE running buffer (50 mM MOPS, 50 mM Tris Base, 0.1% SDS, 1 mM EDTA, pH 7.7).
9. Spin X columns Corning® Costar® Spin-X® centrifuge tube filters cellulose acetate membrane, pore size 0.45 $\mu$m, sterile.
10. SimplyBlue™ SafeStain (Invitrogen)
11. Triethylammonium bicarbonate (Sigma)
12. Sequencing grade modified trypsin (Promega)

**3.1 Transfection and establishing a GFP stable cell line**

1. Plate ~1 x $10^6$ (~60% confluent) in a 10-cm dish in complete medium lacking zeocin/blasticidin. Place cells in a 37°C, 5% $CO_2$ incubator overnight.

2. Pre-warm 1 ml of Opti-MEM serum free medium (Life Technologies) in a 15-ml conical tube.
3. Add 9 μg pOG44 plasmid and 1 μg of pCDNA5/FRT/TO-GFP vector (Note: maintain a 9:1 ratio of pOG44 with pCDNA5/FRT/TO-GFP when scaling target cell dish size).
4. Add 25 μl of polyethylenemine (1 mg/ml) (Polysciences) or other transfection reagent to transfection medium and mix by tapping or vortexing for 15 s
5. Leave the transfection mix for 15 min at room temperature.
6. Add the entire transfection mixture to the target cell dish in a drop-wise fashion and return cells to the incubator.
7. 24 h post-transfection split cells into 3 dishes and maintain in DMEM, 10% FBS, L-glutamine, penicillin and streptomycin for 24 h.
8. Exchange medium with fresh medium containing 15 μg/ml blasticidin and 100 μg/ml hygromycin. Hygromycin selection will cause the death of untransfected cells lacking the integration of pCDNA5/FRT/TO/GFP vector into the FRT site. Replace medium every three days. As a negative control, use the selection medium in untransfected control cells.
9. Check cells daily until majority of cells are dead and a number of colonies have formed. Then begin the expansion of cells for protein expression analysis.
10. Test doxycycline (20 ng/ml) or tetracycline (1 μg/ml) induction time course (Figure 1) for GFP-expression. Fluorescence microscopy can also be used to ensure that every cell expresses GFP upon induction with doxycycline or tetracycline.

**3.2 Purification of GFP-interacting proteins from stable cell lines using DSP crosslinking agent**

1. Induce GFP protein expression with 20 ng/ml doxycycline for a defined time.
2. Wash cells once in ice-cold PBS. Gently add PBS to prevent dislodging cells during washing steps.
3. Lyse by scraping the desired number and size (10cm-15cm) of semi-confluent dishes in complete lysis buffer (above) with 2.5 mg/ml DSP. Transfer lysates onto a falcon tube.
4. Keep the lysates on ice for 30 minutes, then quench crosslinking reaction by adding 6.25 ml of 1M Tris-HCl, pH 7.4 per 25ml lysis buffer (200 mM Tris-HCl final) and incubate for further 30 minutes. (Lysates can be snap frozen in liquid nitrogen and stored at -80 $^{o}$C this stage)
5. Spin lysates at 10,000 x g for 30 min at 4 $^{o}$C. Collect a small aliquot of cleared lysate for immunoblot analysis.
6. Filter the lysates through a filter column to clear the lysate of any debris.
7. To clarified lysate, add 50 μl packed sepharose or agarose beads for 1h at 4 $^{o}$C. This pre-clearing step minimizes non-specific interactions to solid-phase.
8. To the cleared lysate (10-60 mg total protein), add 50 μl of equilibrated GFP-rap beads and incubate for 4 h on a roller at 4$^{o}$C.
9. Wash beads in 4 x 10 ml of complete lysis buffer containing 0.5M NaCl. Addition of NaCl removes weak or low affinity spurious interactions.
10. Wash beads with 10 ml of 50 mM Tris-HCl, pH 7.5, 0.1mM EGTA.
11. Reconstitute beads in 60 μl of 1X LDS sample buffer with 0.1M DTT final, incubate for 1 h at 37 $^{o}$C, boil for 5 min at 95 $^{o}$C. DTT reduces disulfide bond from DSP to remove cross-linking of proteins.

12. Isolate denatured proteins from GFP-trap beads by spinning through spin X columns.

**3.3 Gel electrophoresis of GFP purified protein complexes**

1. Prepare a 10% NuPAGE Bis-Tris (1.5 mm thick) gel for electrophoretic separation of proteins. Rinse wells thoroughly with MOPS running buffer.
2. Load equal amounts of control GFP and GFP-protein of interest into alternating wells (leave empty wells between samples to minimize potential cross contamination during loading).
3. Resolve proteins at ~160 volts for ~45 minutes.
4. Stain using mass spectrometry grade staining reagent containing Coomassie blue dye or Simply Blue.
5. Image gel in clean petri dish, avoid unnecessary handling to minimize keratin contamination.

**4.1 In gel Trypsin digestion**

1. Excise bands of interest for each sample lane (See example Figure 4A).
2. Cut bands into small 1-2 mm cubes with a clean scalpel. Place into clean 1.5 ml centrifuge tubes, minimize keratin contamination by changing gloves frequently.
3. Wash gel bands sequentially in 500 μl of 1) water, 2) 50% acetonitrile, 3) 50 mM $NH_4HCO_3$ (ammomium bicarbonate; prepared fresh) and 4) 50% acetonitrile/50 mM $NH_4HCO_3$. Place samples on shaker for 10 min at room temperature for each wash step. Aspirate wash solution after each step.
4. In-gel alkylation and reduction steps:

a. To each tube add 75 μl of freshly prepared 10 mM DTT/50 mM $NH_4HCO_3$ and incubate for 20 minutes at 37°C. Remove solution and proceed to alkylation step.

b. Add 75 μl of freshly prepared 100 mM iodoacetamide/50 mM $NH_4HCO_3$ and incubate in dark at room temperature for 20 min.

6. Remove liquid and wash bands in 50 mM $NH_4HCO_3$ for 10 min at room temperature.
7. Remove liquid and replace with 50% acetonitrile/50 mM $NH_4HCO_3$ to remove any dye stain. Repeat wash steps until gel pieces are colorless.
8. Add 300 μl of acetonitrile to each tube for 10-15 minutes. Speed vacuum away additional acetonitrile, dry gels pieces will appear opaque.
9. Re-swell gel pieces in 25 mM triethylammonium bicarbonate containing 5 μg/ml of sequencing grade modified porcine trypsin. Place on shaker at 30 °C. After 20-30 minutes, check gel pieces are adequately soaked in solution. If not, add more triethylammonium bicarbonate to assure coverage of bands. Digest gel bands overnight.
10. Following overnight digest, add an equivalent volume of acetonitrile to each gel piece and shake for 10-15 minutes.
11. Transfer supernatant onto a clean tube. Place directly into vacuum centrifuge and spin until dry.
12. To gel pieces, add additional 100 μl of 50% acetonitrile/2.5% formic acid and shake. Add supernatant to dried tubes from step 11. Transfer tubes to vacuum centrifuge until dry. Samples are now ready for 1-dimensional HPLC electrospray ionization. Samples can be stored at -80 °C until ready to perform mass spectrometry analysis.

## 4.2 Mass spectrometry

1. Solubilize digested peptide samples into a working volume of HPLC grade 5% acetonitrile/0.1% formic acid for injection into the HPLC autosampler, typically 20-30 µl volumes are reasonable.
2. Resulting peptides are injected onto a 3-µm $C_{18}$ reversed-phase (RP) resin column (Acclaim PepMap100, 75 µm × 15 cm). Peptides are eluted into a Thermo LTQ Orbitrap over a 40-minute gradient of 2% acetonitrile/0.1% formic acid to 50% of a 90% acetonitrile/0.08% formic acid solvent.
3. Spectra are collected under data-dependent analysis selecting for the 5 most intense MS ions to be fragmented for MS-MS sequencing (See Figure 4B for example). Depending on instrumentation, additional ion fragmentation can be integrated into the data-dependent analysis. Exclusion lists for repeat ions can be developed to increase spectral analysis.
4. Raw data files/spectra can be submitted to in-house proprietary (Mascot, Sequest Sorcerer) or non-proprietary (Global proteome machine analysis www.gpm.org) database search softwares. Once spectra have been assigned and proteins identified, the lists of interacting proteins can be interrogated using ontological and cluster analysis softwares (see section 4.3). Quantitative analysis can be performed using Scaffold, a statistical package for determination of data quality and spectral assignments (*6*).

## 4.3 Bioinformatics analysis of dataset

Analysis of shotgun proteomics lists can be conducted using proprietary software such as Ingenuity Pathways Analysis or public databases such as the Database for Annotation, Visualization and Integrated Discovery (DAVID,

http://david.abcc.ncifcrf.gov) (*7, 8*). These software packages allow for delving into the dataset and assigning molecular function and disease correlation along with developing various other ontological connections within the dataset. Gene Ontology (geneontology.org) can also be used to interrogate and determine term enrichment within a dataset (*9*). Cell localization, function and disease associations can be assigned to groups of proteins to better understand characteristics of the proteome cohort (*10*). Solid phase interactors (i.e. GFP binders) can be compared to the CRAPome database of non-specific interactors as well (crapome.org) (*5*). This repository is useful towards determining ruling out non-specific interactions.

**Notes and Conclusions:**

Affinity purification of target protein complexes provides a convenient way to identify novel binding partners involved in biological processes. Inclusion of GFP tags allows for determination of transfection efficiency by visualization. One caveat is the effect of the large tag on native interactions. Considering this caveat, a strategy to tag target proteins either at the N or C terminus or alternatively at an internal site could provide insight into the effects of GFP fusion on native protein interactions. Careful determination of the induction time for the GFP fusion protein should be considered for potential spurious interactions due to heightened overexpression and effects of high abundance peptides masking identification of lower abundance but biologically important interactions. Using another method of fractionation can improve proteome identification and minimize dampening of signal for low abundance peptides (i.e. strong cation or strong anion HPLC fractionation of peptides or size exclusion chromatography of proteins prior to tryptic digestion). The identification of protein complexes can yield a great deal of information in terms of protein function and involvement in biological processes. The above protocol provides a detailed approach towards conducting protein network characterization using mass spectrometry based technologies. The inclusion of a list of human proteins that interact with GFP under the experimental conditions described above will allow researchers to exclude these as contaminants in similar proteomic approach looking at potential interactors of GFP-tagged proteins (Table 1).

We have employed the above methodology routinely for many proteomic studies (*11-14*) and continue to do so. Affinity purification of protein complexes and analysis by mass-spectrometry is a widely used technique and so variations to the above methodology also exist.

Acknowledgements:

We thank Kirsten McLeod and Janis Stark for help with tissue culture, Thomas Macartney for cloning pCDNA5/FRT/TO/GFP, and the staff at the Sequencing Service (School of Life Sciences, University of Dundee, Scotland) for DNA sequencing. We thank Dave Campbell and the mass-spectrometric team at the MRC Protein Phosphorylation and Ubiquitylation Unit for optimization of protocols relating to sample processing for mass-spectrometry. TDC is supported by the UK MRC Career Development Fellowship. GS is supported by the UK Medical Research Council and the pharmaceutical companies supporting the DSTT (AstraZeneca, Boehringer-Ingelheim, GlaxoSmithKline, Merck-Serono, Pfizer and Johnson & Johnson).

**References:**

(1) Huang, B.X., and Kim, H.Y. (2013) Effective identification of Akt interacting proteins by two-step chemical crosslinking, co-immunoprecipitation and mass spectrometry. *PloS one*. **8**, e61430

(2) Gao, M., McCluskey, P., Loganathan, S.N., and Arkov, A.L. (2014) An in vivo crosslinking approach to isolate protein complexes from Drosophila embryos. *Journal of visualized experiments : JoVE*.

(3) Kim, K.M., Yi, E.C., and Kim, Y. (2012) Mapping protein receptor-ligand interactions via in vivo chemical crosslinking, affinity purification, and differential mass spectrometry. *Methods*. **56**, 161-165

(4) Corgiat, B.A., Nordman, J.C., and Kabbani, N. (2014) Chemical crosslinkers enhance detection of receptor interactomes. *Frontiers in pharmacology*. **4**, 171

(5) Mellacheruvu, D., Wright, Z., Couzens, A.L., Lambert, J.P., St-Denis, N.A., Li, T., Miteva, Y.V., Hauri, S., Sardiu, M.E., Low, T.Y., Halim, V.A., Bagshaw, R.D., Hubner, N.C., Al-Hakim, A., Bouchard, A., Faubert, D., Fermin, D., Dunham, W.H., Goudreault, M., Lin, Z.Y., Badillo, B.G., Pawson, T., Durocher, D., Coulombe, B., Aebersold, R., Superti-Furga, G., Colinge, J., Heck, A.J., Choi, H., Gstaiger, M., Mohammed, S., Cristea, I.M., Bennett, K.L., Washburn, M.P., Raught, B., Ewing, R.M., Gingras, A.C., and Nesvizhskii, A.I. (2013) The CRAPome: a contaminant repository for affinity purification-mass spectrometry data. *Nature methods*. **10**, 730-736

(6) Searle, B.C. (2010) Scaffold: a bioinformatic tool for validating MS/MS-based proteomic studies. *Proteomics*. **10**, 1265-1269

(7) Yang, L.J., Ma, D.Q., and Cui, H. (2014) Proteomic analysis of immature rat pups brain in response to hypoxia and ischemia challenge. *International journal of clinical and experimental pathology*. **7**, 4645-4660

(8) Huang da, W., Sherman, B.T., and Lempicki, R.A. (2009) Systematic and integrative analysis of large gene lists using DAVID bioinformatics resources. *Nature protocols*. **4**, 44-57

(9) Carnielli, C.M., Winck, F.V., and Paes Leme, A.F. (2015) Functional annotation and biological interpretation of proteomics data. *Biochimica et biophysica acta*. **1854**, 46-54

(10) Pawar, H., Kashyap, M.K., Sahasrabuddhe, N.A., Renuse, S., Harsha, H.C., Kumar, P., Sharma, J., Kandasamy, K., Marimuthu, A., Nair, B., Rajagopalan, S., Maharudraiah, J., Premalatha, C.S., Kumar, K.V., Vijayakumar, M., Chaerkady, R., Prasad, T.S., Kumar, R.V., Kumar, R.V., and Pandey, A. (2011) Quantitative tissue proteomics of esophageal squamous cell carcinoma for novel biomarker discovery. *Cancer biology & therapy*. **12**, 510-522

(11) Vogt, J., Dingwell, K.S., Herhaus, L., Gourlay, R., Macartney, T., Campbell, D., Smith, J.C., and Sapkota, G.P. (2014) Protein associated with SMAD1 (PAWS1/FAM83G) is a substrate for type I bone morphogenetic protein receptors and modulates bone morphogenetic protein signalling. *Open biology*. **4**, 130210
(12) Al-Salihi, M.A., Herhaus, L., Macartney, T., and Sapkota, G.P. (2012) USP11 augments TGFbeta signalling by deubiquitylating ALK5. *Open biology*. **2**, 120063
(13) Herhaus, L., Al-Salihi, M., Macartney, T., Weidlich, S., and Sapkota, G.P. (2013) OTUB1 enhances TGFbeta signalling by inhibiting the ubiquitylation and degradation of active SMAD2/3. *Nature communications*. **4**, 2519
(14) Herhaus, L., Al-Salihi, M.A., Dingwell, K.S., Cummins, T.D., Wasmus, L., Vogt, J., Ewan, R., Bruce, D., Macartney, T., Weidlich, S., Smith, J.C., and Sapkota, G.P. (2014) USP15 targets ALK3/BMPR1A for deubiquitylation to enhance bone morphogenetic protein signalling. *Open biology*. **4**, 140065

## Figure Legends

**Figure 1.** A time course of doxycycline induction of GFP-expression in T-Rex 293 GFP cells. Establishing a time course of robust induction of the target protein to determine appropriate amount of protein production for interaction analysis. Here 6-16 hours of doxycycline induction is appropriate to perform protein isolation and downstream mass spectrometry analysis.

**Figure 2.** Dithiobis (succinimidyl propionate): In blue brackets indicates carbodiimide that reacts with primary amines releasing the NHS ring while cross-linking polypeptides, red disulfide bond indicates bond that reacts with reducing reagent to cleave linkage and release cross-linked proteins or polypeptides.

**Figure 3.** Schematic work-flow for GFP-trap purification and mass spectrometry analysis of co-purified proteins.

**Figure 4.** A. Affinity-purified GFP was resolved by SDS-PAGE electrophoeresis with molecular weight markers (kDa) indicated in the first lane. Brackets indicate one excised gel-piece, which was tryptically digested for mass spectrometric electrospray analysis. B. Identification of a prominent protein from the 100 kDa excised band in GFP lane. GFP interactors are compared to target protein of interest to exclude non-specific GFP or bead interacting proteins Included in Table 1 are highly enriched promiscuous GFP contaminants that can be excluded from target protein pulldown analysis.

**Table 1**. GFP-binding proteins in a DSP cross-linked pulldown experiment. Proteins are rank ordered by normalized spectral count enrichment using Scaffold software (v.4.2.1).

FIGURE 1.

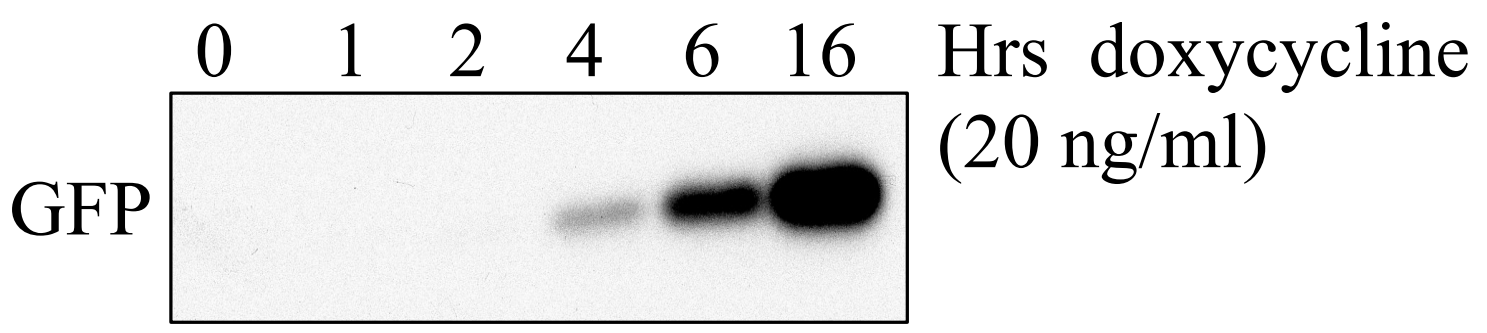

FIGURE 2.

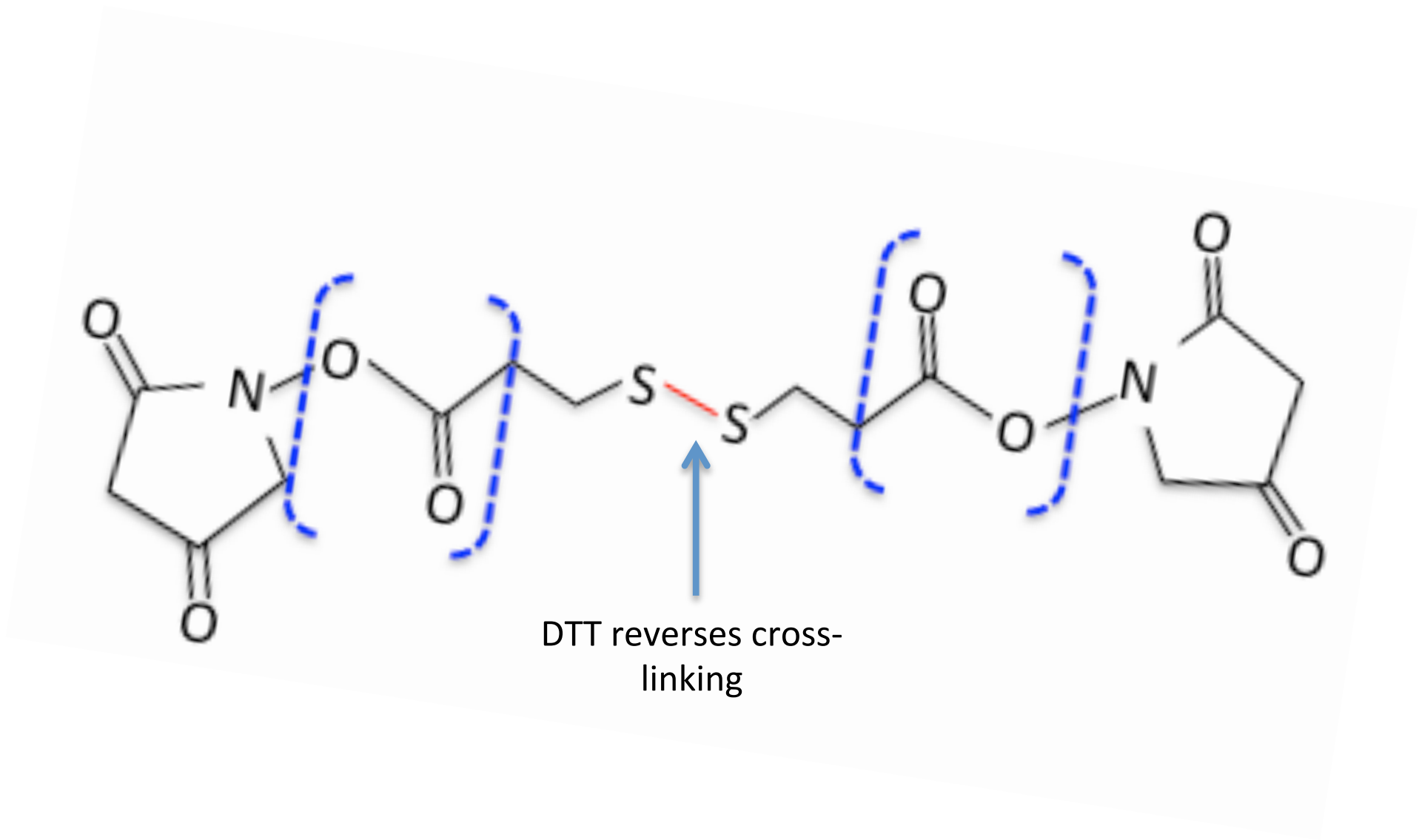

FIGURE 3.

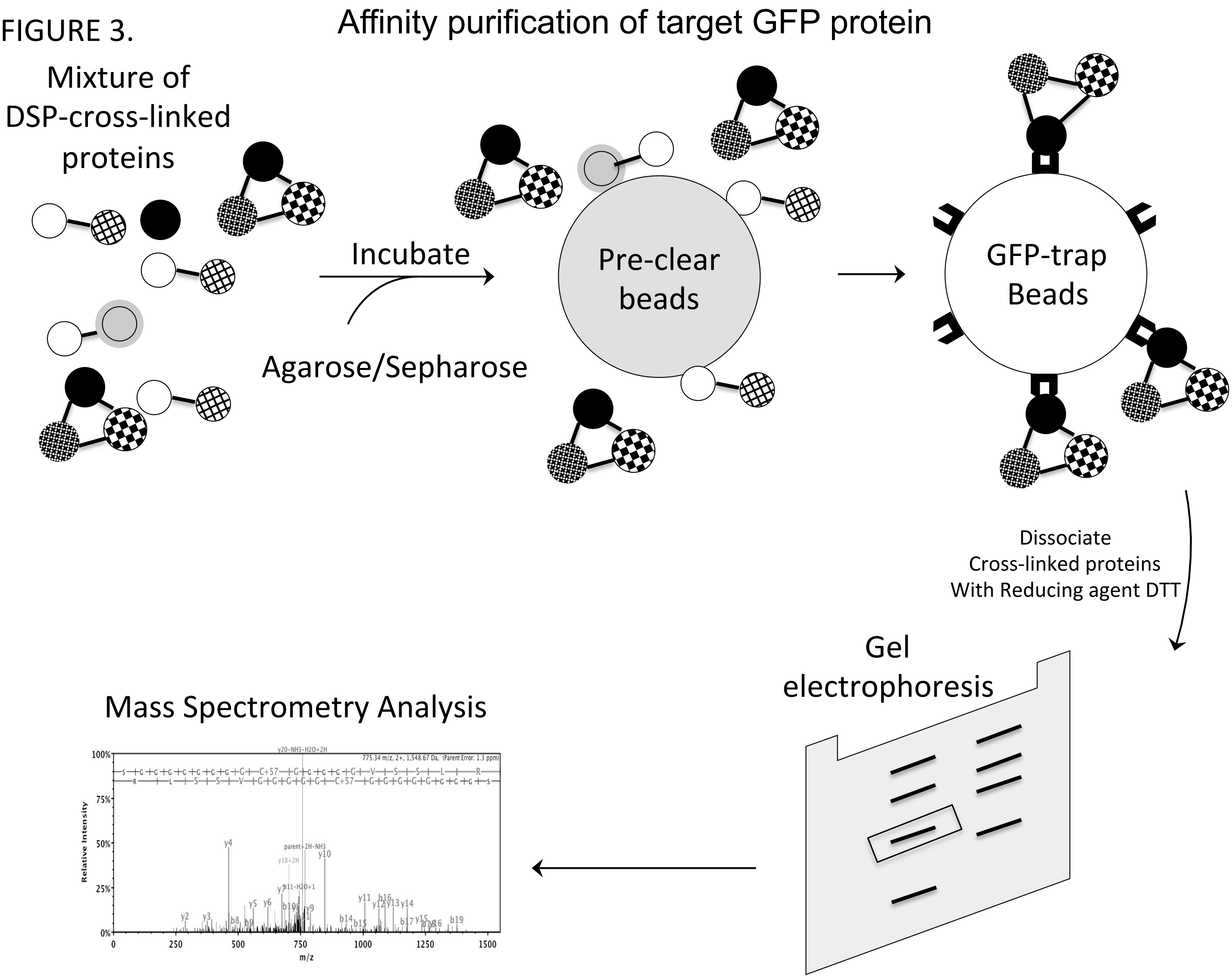
Affinity purification of target GFP protein
Mixture of DSP-cross-linked proteins
Incubate
Agarose/Sepharose
Pre-clear beads
GFP-trap Beads
Dissociate
Cross-linked proteins
With Reducing agent DTT
Gel electrophoresis
Mass Spectrometry Analysis
Relative Intensity
m/z

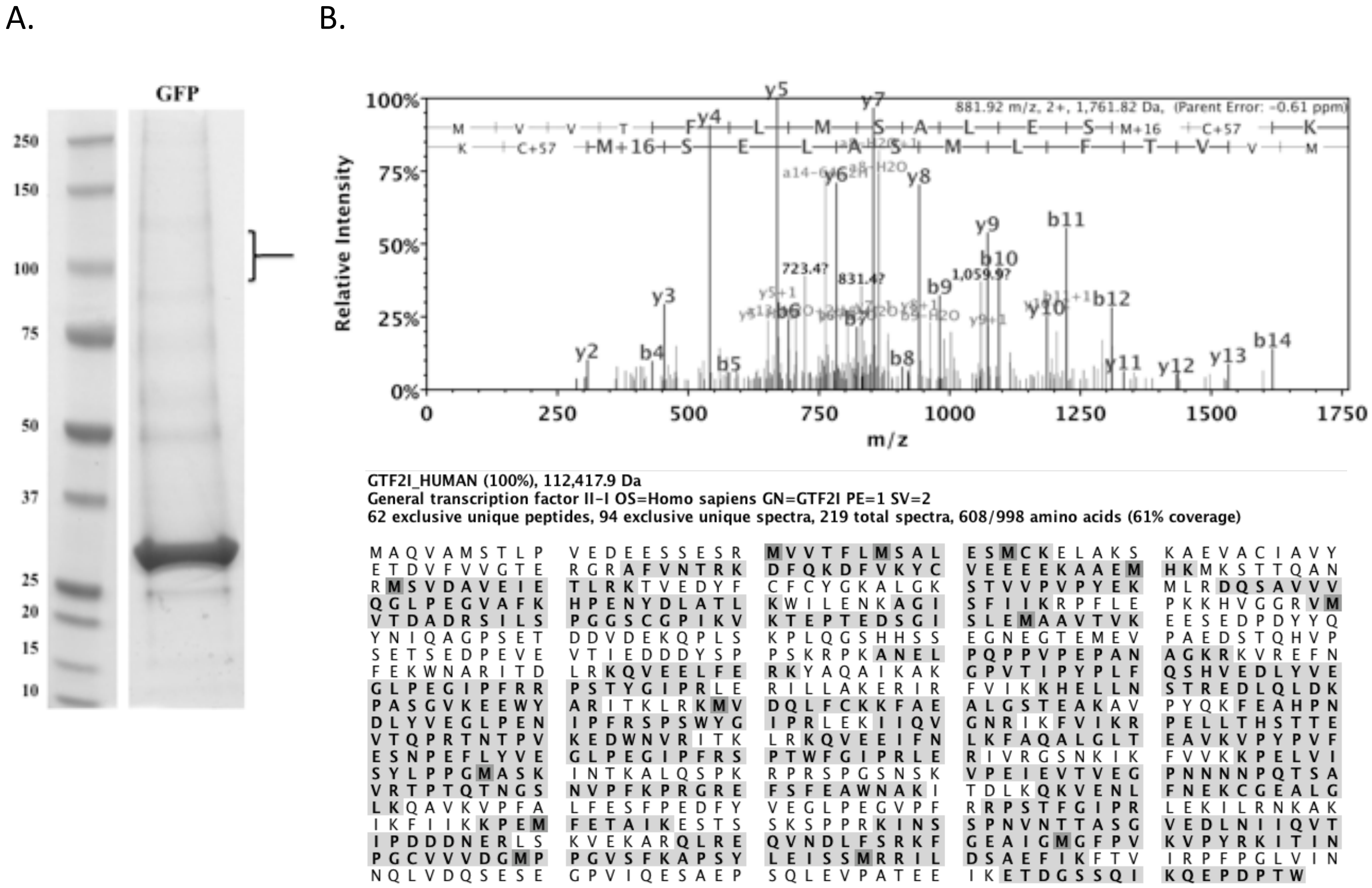
A.
B.
GFP
250
150
100
75
50
37
25
20
15
10
881.92 m/z, 2+, 1,761.82 Da, (Parent Error: -0.61 ppm)
100%
75%
50%
25%
0%
Relative Intensity
0
250
500
750
1000
1250
1500
1750
m/z
y2
y3
y4
y5
y6
y7
y8
y9
y10
y11
y12
y13
b4
b5
b6
b7
b8
b9
b10
b11
b12
b14
723.4?
831.4?
1,059.9?
GTF2I_HUMAN (100%), 112,417.9 Da
General transcription factor II-I OS=Homo sapiens GN=GTF2I PE=1 SV=2
62 exclusive unique peptides, 94 exclusive unique spectra, 219 total spectra, 608/998 amino acids (61% coverage)
MAQVAMSTLP VEDEESSESR MVVTFLMSAL ESMCKELAKS KAEVACIAVY
ETDVFVVGTE RGRAFVNTRK DFQKDFVKYC VEEEEKAAEM HKMKSTTQAN
RMSVDAVEIE TLRKTVEDYF CFCYGKALGK STVVPVPYEK MLRDQSAVVV
QGLPEGVAFK HPENYDLATL KWILENKAGI SFIIKRPFLE PKKHVGGRVM
VTDADRSILS PGGSCGPIKV KTEPTEDSGI SLEMAAVTVK EESEDPDYYQ
YNIQAGPSET DDVDEKQPLS KPLQGSHHSS EGNEGTEMEV PAEDSTQHVP
SETSEDPEVE VTIEDDDYSP PSKRPKANEL PQPPVPEPAN AGKRKVREFN
FEKWNARITD LRKQVEELFE RKYAQAIKAK GPVTIPYPLF QSHVEDLYVE
GLPEGIPFRR PSTYGIPRLE RILLAKERIR FVIKKHELLN STREDLQLDK
PASGVKEEWY ARITKLRKMV DQLFCKKFAE ALGSTEAKAV PYQKFEAHPN
DLYVEGLPEN IPFRSPSWYG IPRLEKIIQV GNRIKFVIKR PELLTHSTTE
VTQPRTNTPV KEDWNVRITK LRKQVEEIFN LKFAQALGLT EAVKVPYPVF
ESNPEFLYVE GLPEGIPFRS PTWFGIPRLE RIVRGSNKIK FVVKKPELVI
SYLPPGMASK INTKALQSPK RPRSPGSNSK VPEIEVTVEG PNNNNPQTSA
VRTPTQTNGS NVPFKPRGRE FSFEAWNAKI TDLKQKVENL FNEKCGEALG
LKQAVKVPFA LFESFPEDFY VEGLPEGVPF RRPSTFGIPR LEKILRNKAK
IKFIIKKPEM FETAIKESTS SKSPPRKINS SPNVNTTASG VEDLNIIQVT
IPDDDNERLS KVEKARQLRE QVNDLFSRKF GEAIGMGFPV KVPYRKITIN
PGCVVVDGMP PGVSFKAPSY LEISSMRRIL DSAEFIKFTV IRPFPGLVIN
NQLVDQSESE GPVIQESAEP SQLEVPATEE IKETDGSSQI KQEPDPTW

FIGURE 4.

Table 1.

| Gene Symbol | Spectral Count (Normalized ) | MW (kDa) |
|---|---|---|
| ACTN4 | 233.77 | 105 |
| EPPK1 | 204.55 | 556 |
| PLEC | 204.55 | 532 |
| GTF2I | 175.33 | 112 |
| PRDX1 | 146.11 | 22 |
| YWHAZ | 146.11 | 28 |
| DSG1 | 146.11 | 114 |
| EEF1A1 | 116.89 | 50 |
| YWHAE | 116.89 | 29 |
| GRP78 | 116.89 | 72 |
| RPS27A | 116.89 | 18 |
| UBA52 | 116.89 | 15 |
| ENO1 | 116.89 | 47 |
| GNB2L1 | 116.89 | 35 |
| ALB | 116.89 | 69 |
| KRT77 | 116.89 | 62 |
| EEF1A2 | 87.664 | 50 |
| ACTN1 | 87.664 | 103 |
| TUFM | 87.664 | 50 |
| RPS3 | 87.664 | 27 |
| NONO | 87.664 | 54 |
| SF3B3 | 87.664 | 136 |
| PCMT1 | 87.664 | 25 |
| ACLY | 87.664 | 121 |
| VCL | 87.664 | 124 |
| PDIA6 | 87.664 | 48 |
| GRP94 | 87.664 | 92 |
| MCM6 | 87.664 | 93 |
| NUP98 | 87.664 | 198 |
| PRDX4 | 87.664 | 31 |
| PCCB | 87.664 | 58 |
| TPI1 | 87.664 | 31 |
| HNRNPUL1 | 87.664 | 96 |
| PDAP1 | 87.664 | 21 |
| CLTC | 58.443 | 192 |
| ACTG1 | 58.443 | 42 |
| SPTAN1 | 58.443 | 285 |
| EPRS | 58.443 | 171 |
| SPTBN1 | 58.443 | 275 |
| NUP214 | 58.443 | 214 |
| RANBP2 | 58.443 | 358 |
| HNRNPM | 58.443 | 78 |
| GART | 58.443 | 108 |
| HSPD1 | 58.443 | 61 |
| NUP155 | 58.443 | 155 |
| PRDX2 | 58.443 | 22 |
| MTHFD1 | 58.443 | 102 |
| HNRNPU | 58.443 | 91 |
| CCT4 | 58.443 | 58 |
| TCP1 | 58.443 | 60 |
| HNRNPA1 | 58.443 | 39 |
| YWHAH | 58.443 | 28 |
| KHSRP | 58.443 | 73 |
| SND1 | 58.443 | 102 |
| HNRNPA2B1 | 58.443 | 37 |
| MARS | 58.443 | 101 |
| NUP93 | 58.443 | 93 |
| NUMA1 | 58.443 | 238 |
| PHGDH | 58.443 | 57 |
| GDI1 | 58.443 | 51 |
| YWHAG | 58.443 | 28 |
| PSMA7 | 58.443 | 28 |
| TPM3 | 58.443 | 33 |
| USP5 | 58.443 | 96 |
| NUP107 | 58.443 | 106 |
| RAE1 | 58.443 | 41 |
| PRPF6 | 58.443 | 107 |
| TAGLN2 | 58.443 | 22 |
| RPL11 | 58.443 | 20 |
| SDHB | 58.443 | ? |
| AARS | 58.443 | 107 |
| HIST1H4A | 58.443 | 11 |
| RAB5C | 58.443 | 23 |
| MIF | 58.443 | 12 |
| MOV10 | 58.443 | 114 |
| UHRF1 | 58.443 | 90 |
| FUBP1 | 58.443 | 68 |
| ACTC1 | 29.221 | 42 |
| ACTA1 | 29.221 | 42 |
| ACTBL2 | 29.221 | 42 |
| POTEE | 29.221 | 121 |
| USP9X | 29.221 | 292 |
| CSNK1D | 29.221 | 47 |
| CNOT1 | 29.221 | 267 |
| CCT2 | 29.221 | 57 |
| ATP5A1 | 29.221 | 60 |
| CCT8 | 29.221 | 60 |
| EIF4A1 | 29.221 | 46 |
| EIF4A2 | 29.221 | 46 |
| MCM3 | 29.221 | 91 |
| MCM7 | 29.221 | 81 |
| DDB1 | 29.221 | 127 |
| PABPC4 | 29.221 | 71 |
| PABPC1 | 29.221 | 71 |
| NUP205 | 29.221 | 228 |
| SMC3 | 29.221 | 142 |
| PARP1 | 29.221 | 113 |
| NUP188 | 29.221 | 196 |
| PSMD1 | 29.221 | 106 |
| RAD50 | 29.221 | 154 |
| GAPDH | 29.221 | 36 |
| PSMC5 | 29.221 | 46 |
| MCM4 | 29.221 | 97 |
| VCP | 29.221 | 89 |
| LARS | 29.221 | 134 |
| IKBKAP | 29.221 | 150 |
| DARS | 29.221 | 57 |
| SFPQ | 29.221 | 76 |
| SFPQ | 29.221 | 76 |
| AHCY | 29.221 | 48 |
| SMC4 | 29.221 | 147 |
| KRT18 | 29.221 | 48 |
| DDX5 | 29.221 | 69 |
| MYO6 | 29.221 | 150 |
| UBA1 | 29.221 | 118 |
| YWHAQ | 29.221 | 28 |
| SDHA | 29.221 | ? |
| RPS4X | 29.221 | 30 |
| DNAJC7 | 29.221 | 56 |
| CSDE1 | 29.221 | 89 |
| UBAP2L | 29.221 | 115 |
| LDHB | 29.221 | 37 |
| NUP88 | 29.221 | 84 |
| RPSA | 29.221 | 33 |
| CDC5L | 29.221 | 92 |
| PCNA | 29.221 | 29 |
| UPF1 | 29.221 | 124 |
| GDI2 | 29.221 | 51 |
| COPB2 | 29.221 | 102 |
| PHB2 | 29.221 | 33 |
| HDLBP | 29.221 | 141 |
| PDIA3 | 29.221 | 57 |
| ADAR | 29.221 | 136 |
| RTCB | 29.221 | 55 |
| ADE2/PAICS | 29.221 | 47 |
| ALDH18A1 | 29.221 | 87 |
| PSMB5 | 29.221 | 28 |
| SHMT2 | 29.221 | 56 |
| DDX1 | 29.221 | 82 |
| FKBP4 | 29.221 | 52 |
| RPS3A | 29.221 | 30 |
| MST4 | 29.221 | 47 |
| FAM115A | 29.221 | 102 |
| OGT1 | 29.221 | 117 |
| HNRNPH1 | 29.221 | 49 |
| TPM4 | 29.221 | 29 |
| NPEPPS | 29.221 | 103 |
| KIF11 | 29.221 | 119 |
| RPL4 | 29.221 | 48 |
| PUM1 | 29.221 | 126 |
| TARS | 29.221 | 83 |
| ALDOA | 29.221 | 39 |
| RAN | 29.221 | 24 |
| RPS11 | 29.221 | 18 |
| DIAPH1 | 29.221 | 141 |
| HNRNPA3 | 29.221 | 40 |
| NAP1L1 | 29.221 | 45 |
| BUB3 | 29.221 | 37 |
| PDCD6IP | 29.221 | 96 |
| RPL10A | 29.221 | 25 |
| RPS8 | 29.221 | 24 |
| LDHA | 29.221 | 37 |

# Table 1. Cont'd

| Gene Symbol | Spectral Count (Normalized ) | MW (kDa) |
|---|---|---|
| ABCE1 | 29.221 | 67 |
| ILF3 | 29.221 | 95 |
| DNAJC10 | 29.221 | 91 |
| KRT31 | 29.221 | 47 |
| KRT35 | 29.221 | 50 |
| KRT33B | 29.221 | 46 |
| KRT33A | 29.221 | 46 |
| C14orf166 | 29.221 | 28 |
| HSD17B10 | 29.221 | 27 |
| RPL17 | 29.221 | 21 |
| SERPINB4 | 29.221 | 45 |
| PRDX6 | 29.221 | 25 |
| SERPINB6 | 29.221 | 43 |
| SARS2 | 29.221 | 58 |
| EPS15 | 29.221 | 99 |
| HNRNPH2 | 29.221 | 49 |
| DPYSL2 | 29.221 | 62 |
| PRDX3 | 29.221 | 28 |
| DDX42 | 29.221 | 103 |
| TBCE | 29.221 | 59 |
| CSRP2 | 29.221 | 21 |
| CKAP4 | 29.221 | 66 |
| RBM39 | 29.221 | 59 |
| EIF6 | 29.221 | 27 |
| PMPCB | 29.221 | 54 |
| RPL13A | 29.221 | 24 |
| CSTF1 | 29.221 | 48 |
| PPA1 | 29.221 | 33 |
| MICAL1 | 29.221 | 118 |
| TFAM | 29.221 | 29 |
| SEC22B | 29.221 | 25 |
| JUP | 29.221 | 82 |
| PPIF | 29.221 | 22 |
| TBCB | 29.221 | 27 |
| ST13 | 29.221 | 41 |
| SYNCRIP | 29.221 | 70 |
| CBX3 | 29.221 | 21 |
| RPL19 | 29.221 | 23 |
| RANBP1 | 29.221 | 23 |
| RPL28 | 29.221 | 16 |
| PFN1 | 29.221 | 15 |
| RBM10 | 29.221 | 104 |
| LMNB2 | 29.221 | 68 |
| RPL17 | 29.221 | 21 |
| SERPINB4 | 29.221 | 45 |
| PRDX6 | 29.221 | 25 |
| SERPINB6 | 29.221 | 43 |
| SARS2 | 29.221 | 58 |
| EPS15 | 29.221 | 99 |
| HNRNPH2 | 29.221 | 49 |
| DPYSL2 | 29.221 | 62 |
| PRDX3 | 29.221 | 28 |
| DDX42 | 29.221 | 103 |
| TBCE | 29.221 | 59 |
| CSRP2 | 29.221 | 21 |

| Gene Symbol | Spectral Count (Normalized ) | MW (kDa) |
|---|---|---|
| CKAP4 | 29.221 | 66 |
| RBM39 | 29.221 | 59 |
| EIF6 | 29.221 | 27 |
| PMPCB | 29.221 | 54 |
| RPL13A | 29.221 | 24 |
| CSTF1 | 29.221 | 48 |
| PPA1 | 29.221 | 33 |
| MICAL1 | 29.221 | 118 |
| TFAM | 29.221 | 29 |
| SEC22B | 29.221 | 25 |
| JUP | 29.221 | 82 |
| PPIF | 29.221 | 22 |
| TBCB | 29.221 | 27 |
| ST13 | 29.221 | 41 |
| SYNCRIP | 29.221 | 70 |
| CBX3 | 29.221 | 21 |
| RPL19 | 29.221 | 23 |
| RANBP1 | 29.221 | 23 |
| RPL28 | 29.221 | 16 |
| PFN1 | 29.221 | 15 |
| RBM10 | 29.221 | 104 |
| LMNB2 | 29.221 | 68 |
| HIST1H1E | 29.221 | 22 |
| ACAA2 | 29.221 | 42 |
| HNRNPR | 29.221 | 71 |
| PDE12 | 29.221 | 67 |
| SF3B4 | 29.221 | 44 |
| UBE2N | 29.221 | 17 |
| LIG3 | 29.221 | 113 |
| ERAL1 | 29.221 | 48 |
| CARS | 29.221 | 85 |
| UCHL3 | 29.221 | 26 |
| SCYL2 | 29.221 | 104 |
| CBX1 | 29.221 | 21 |
| ERO1L | 29.221 | 54 |